\documentclass{PoS}

\usepackage{subfig}
\usepackage{lineno}

\title{Detector considerations for a HAWC southern observatory}

\ShortTitle{HAWC South}

\author{Michael DuVernois$^{\dagger}$ for the HAWC Collaboration
  \\
    {\it 
     $\dagger$Department of Physics and Wisconsin IceCube Particle Astrophysics Center
      (WIPAC), University of Wisconsin--Madison, Madison, WI, USA\\
    }
E-mail: \email{michael.duvernois@icecube.wisc.edu}\\
For a complete author list, see \href{http://www.hawc-observatory.org/collaboration/icrc2015.php}{www.hawc-observatory.org/collaboration/icrc2015.php}.
}

\abstract{
 The High-Altitude Water Cherenkov (HAWC) observatory in central Mexico is currently the world's only synoptic survey instrument for gamma rays above 1 TeV. Because there is significant interest in covering the full TeV sky with a survey instrument, we have examined options for a Southern Hemisphere extension to HAWC. In addition to providing all-sky coverage of TeV sources, a southern site could complement existing surveys of the densest part of the Galactic Plane, provide continuous monitoring of Galactic and extragalactic transient sources in both Hemispheres, and simplify the analysis of spatially extended signals such as diffuse gamma rays and the TeV cosmic-ray anisotropy. To take advantage of the air-shower physics and lower the energy threshold of the experiment as much as possible, a high altitude site above 5000 m a.s.l (vs. 4100 m a.s.l. at the current site in Mexico) has been specified. To facilitate efficient detector construction at such altitudes, the detector tanks would be assembled at lower altitude and delivered to the site. An all-digital communications and data acquisition scheme is proposed. Possible designs include taking advantage of digital optical module technology from the IceCube experiment as well as new custom electronics. We discuss the physics potential of such an experiment, focusing on the energy threshold, angular resolution, and background suppression capability of the experiment, as well as the advantages of full-sky coverage above 1 TeV.
   
  \vspace{4mm}
  
{\bf Corresponding authors:}
 \speaker{M.~DuVernois}\\
}

\FullConference{The 34th International Cosmic Ray Conference,\\
  30 July- 6 August, 2015\\
  The Hague, The Netherlands}

\begin{document}

\section{HAWC Observatory}

The HAWC Collaboration has built a high-energy gamma-ray observatory at 19 degrees North latitude and 4100 m a.s.l. in the Sierra Negra--Pico de Orizaba saddle valley in the State of Puebla, Mexico. The array consists of 300 water Cherenkov detector (WCD) tanks instrumented with photomultiplier tubes (PMTs) to detect and record incident gamma and cosmic ray air showers via the Cherenkov emission produced as the shower particles traverse the tank water volume. The full detector was completed and inaugurated in early 2015 with the partial detector taking data during construction.\cite{hawc}

The 300 tanks and 1200 PMTs in HAWC have an active collecting area of about 12,000 m$^{2}$ over a total site area of about 22,000 m$^{2}$. HAWC can detect air showers with energies from about 100 GeV to hundreds of TeV. The direction of the primary particle is reconstructed from the arrival time distribution in the array, the energy from the PMT charge signals, and composition (distinguishing astrophysical photons from the much more abundant hadrons) from the shower topology. Charge 
deposited in the tanks is estimated by a pair of time-over-threshold (ToT) measurements for each PMT.

The primary physics goals of the HAWC Observatory are the detection of new TeV gamma-ray sources and the investigation of transient behaviors at TeV. The observatory complements all-sky lower-energy coverage from the Fermi and SWIFT satellites and narrow-beam observations of the high-energy sky with Imaging Air Cherenkov Telescopes (IACT) such as MAGIC and VERITAS in the Northern Hemisphere and HESS in the Southern Hemisphere. Each day HAWC surveys about 2/3 of the sky from its low latitude location.

The future of gamma-ray astronomy will be significantly different than the present with VERITAS scheduled for shutdown soon and the Cherenkov Telescope Array (CTA) collaboration 
looking to build IACTs in both the Northern and Southern Hemispheres. With HESS already operating and CTA soon to be building in the Southern Hemisphere, a wide field of view instrument capable of acting as a survey instrument for the enhanced deep sensitivity of CTA for the southern sky seems more than sensible. To keep pace with the evolution from VERITAS to CTA, a similar increase in effective area and sensitivity from HAWC in the north (Mexico) to HAWC in the south (HAWC-South) is required.

As an interim step for detector development, and also to enhance the ability of the HAWC array to pinpoint shower core locations when the cores are off the edge of the array, there is a plan to extend
the current HAWC detector using a sparse array of ``outrigger" tanks. The electronics for the outriggers will be a development platform for HAWC-South electronics efforts. The outriggers are described below and are a technological link, in both tank hardware and electronics, from HAWC to a potential larger, higher-altitude HAWC-South.

\section{Outriggers}
With the construction of the 300 HAWC tanks completed, and routine data-taking underway, the HAWC collaboration is exploring ways to increase the sensitivity of the detector. The compact array of tanks in HAWC has a large perimeter/area ratio, and this large amount of 'edge' means that a significant fraction of the high-energy showers that trigger the detector are located outside the main array. In the HAWC-predecessor experiment Milagro, this was mitigated with a thin sampling of small water tanks each with one PMT arrayed outside of the dense inner region. This allows for accurate core and energy measurements even for events which fall somewhat outside of the inner region of the detector. Simulations have shown that $O(100)$ small water tanks (perhaps commercial water storage tanks which are ubiquitous in Mexico) with single PMTs in each could double the effective area
 of the experiment.

\begin{figure}[th]
\vspace{-1.2in}
    \includegraphics[width=\textwidth]{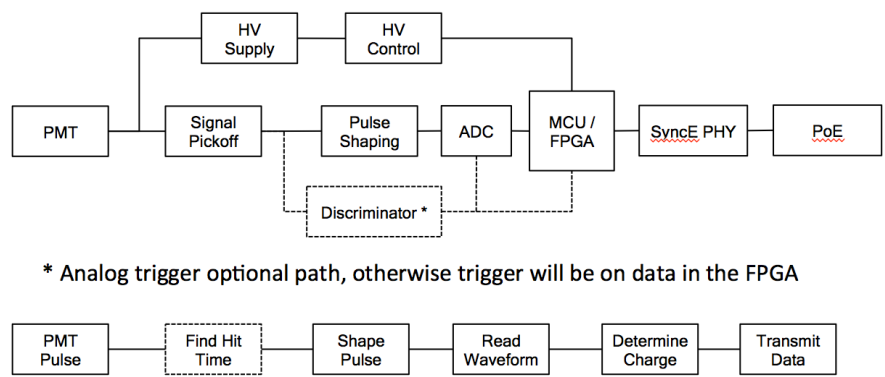}
\vspace{-1.5in}
    \caption{Block diagram for the prototype outrigger electronics.} 
    \label{fig:outrigger}
 \end{figure}

The HAWC detector uses time-over-threshold (ToT) electronics with PMT signals transmitted to a central counting house via analog cables. The electronics were partially inherited from the Milagro experiment and new modules can no longer be produced. In addition, it is not practical to transmit analog signals over the longer distances required in an outrigger array or larger detector. Therefore, distributed electronics are the preferred solution are the preferred solution, and a simple electronics layout is shown in Fig.~\ref{fig:outrigger}. This scheme takes advantage of inexpensive, low-power FADCs and FPGAs to digitize and then generate a
total charge (integral) and time stamp (via synchronous Ethernet) which mimics the data of HAWC, 
but through a very different path. At HAWC, the data rate is between 25--50k hits/s over a dynamic range of 0.25 to 10k photoelectrons which constrains the availability of commercial readout schemes.
The full FADC is available for calibration data, or pre-scaled event 
readouts, while allowing for the simpler (and HAWC-compatible) time plus charge measurement.

\begin{figure}[ht]
 \vspace{-1.5in}
   \includegraphics[width=\textwidth]{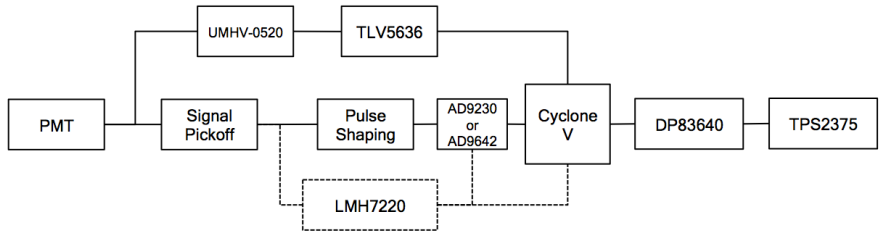}
 \vspace{-1.8in}
    \caption{Chip-level hardware selections for the principle components in the outrigger electronics.} 
    \label{fig:parts}
\end{figure}

Fig.~\ref{fig:parts} shows the same block diagram as Fig.~1 in terms of 
selected, available parts. This set of electronics is undergoing prototyping, with lab versions expected operational in late 2015. The electronics
are equivalent (total charge and arrival time) to the existing system
at the trigger and analysis levels. However, the outrigger electronics system is more flexible, allowing
for full waveforms on request, local histogramming, pre-scaling, and feature extraction. This development effort has been taking advantage of overlapping parts and requirements with the
IceCube Gen2 DOM (\cite{gen2}) development, for example, with common high voltage modules,
shaping electronics, and similar programmable logic.

This set of electronics, and work on small commercial
water tanks, could inform the design of the HAWC-South
detector which, for reasons detailed below, is likely to require smaller tanks that can be transported 
completely outfitted (save for the water) up the mountain, and distributed low-power (possibly solar-powered) electronics.
Full waveform digitization is not ruled out however as the down-stream computing, which would be 
the major limiting factor right now, advances in rough accord with Moore's Law.

\section{HAWC-South}
There are four general areas of improvement in capability for a Southern Hemisphere 
large area water Cherenkov detector that are of interest in moving into the CTA-era of TeV
gamma-ray sensitivities. These gains would be in line with the sensitivity improvements of CTA
over VERITAS, or put another way, a HAWC-South would need to be large enough to act as an
interesting survey instrument for CTA. 

These areas of improvement are higher altitude, larger area, improved hadronic 
rejection, and improved shower sensitivity. Although we'll consider these one at a time, these
issues are all tightly coupled to each other: for example, higher altitude inherently gives more
``ground level" particles for a given primary energy, effectively improving both the shower
detection sensitivity and the total effective area. 

\begin{figure}[h]
\hspace{1.2in}
  \includegraphics[width=.7\textwidth]{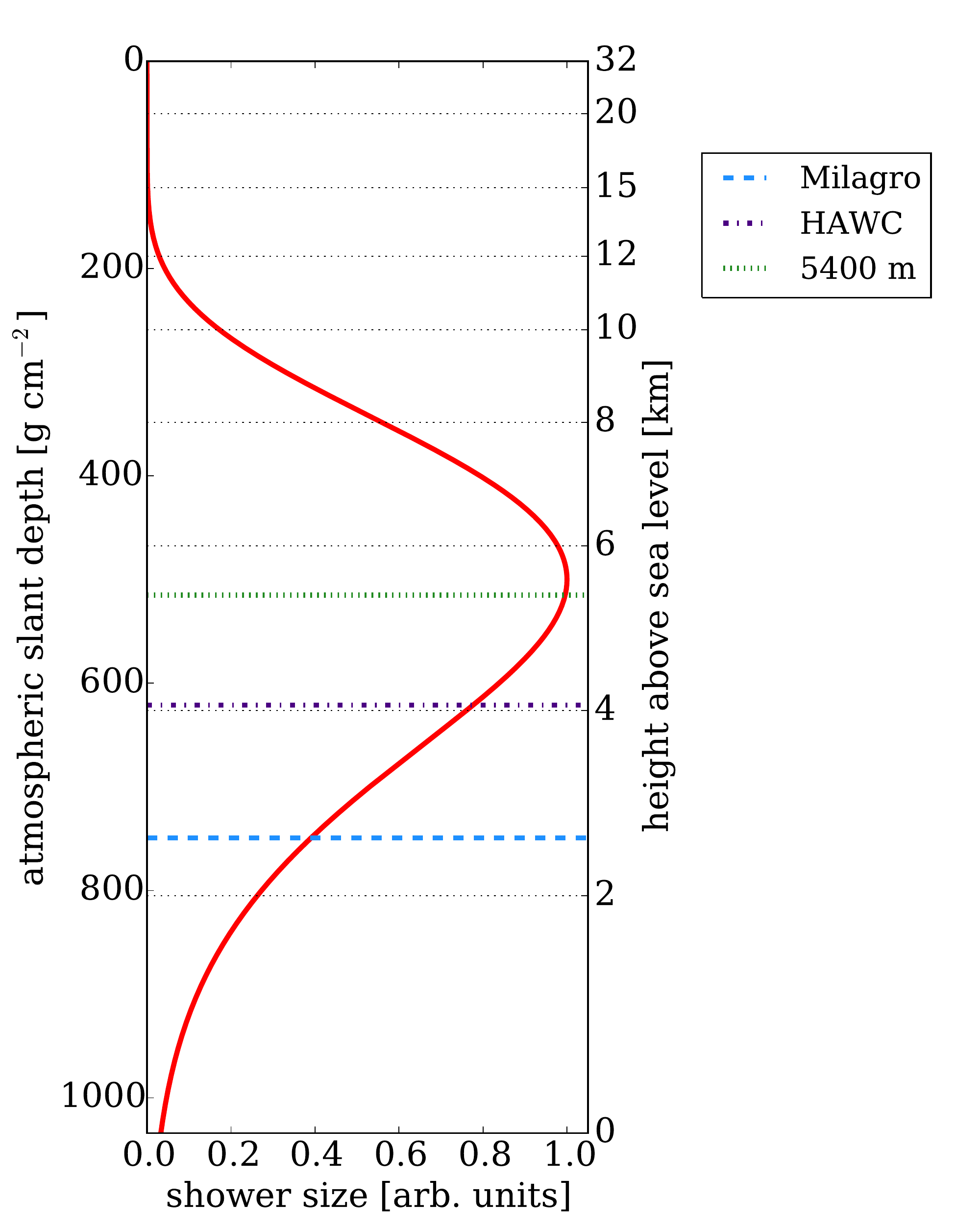}
  \caption{Air shower development with Milagro, HAWC, and potential 5400m site.}
    \label{fig:altitude}
\end{figure}

\subsection{How much higher can you get?}
Increasingly the altitude of the detector site is a huge win in terms of signal strength, lower energy
threshold, and effective
area, but subject to the constraints of both human physiology and land area available at 
a given altitude. At lower energies, the ability to cross-check with IACTs is enhanced.
For a rough feel of the altitudes under consideration and the air shower 
profile, see Fig.~\ref{fig:altitude}. Sites at 5400 m are available in the Chilean Atacama Desert, for example the University of Tokyo Atacama Observatory is being built at 5640 m and a number of
nearby 20,000 m$^2$ hilltops nearby are at altitudes between 5200 and 5400 m. The well-known
cosmic-ray observatory at Chacaltaya in Bolivia is above 5200 m as well. (See \cite{wiki} for a sensible
starting point to look for existing astronomical observatories as potential Southern sites.)

At these altitudes, freezing of the detector water and the inability of unadapted personnel to work
are major problems. Insulation and passive solar gain on the tanks are a potential solution to the former
issue. For site-work, it seems likely that the detector (water tank, PMT, and electronics) would
need to be constructed at lower altitude and then delivered to the high altitude site for later
filling. Roads to the high-altitude Atacama sites appear to be reasonable; see Google Earth's imagery
of the University of Tokyo Atacama Observatory or the ALMA Observatory sites.

\begin{figure}[h]
\hspace{1in}
    \includegraphics[width=.7\textwidth]{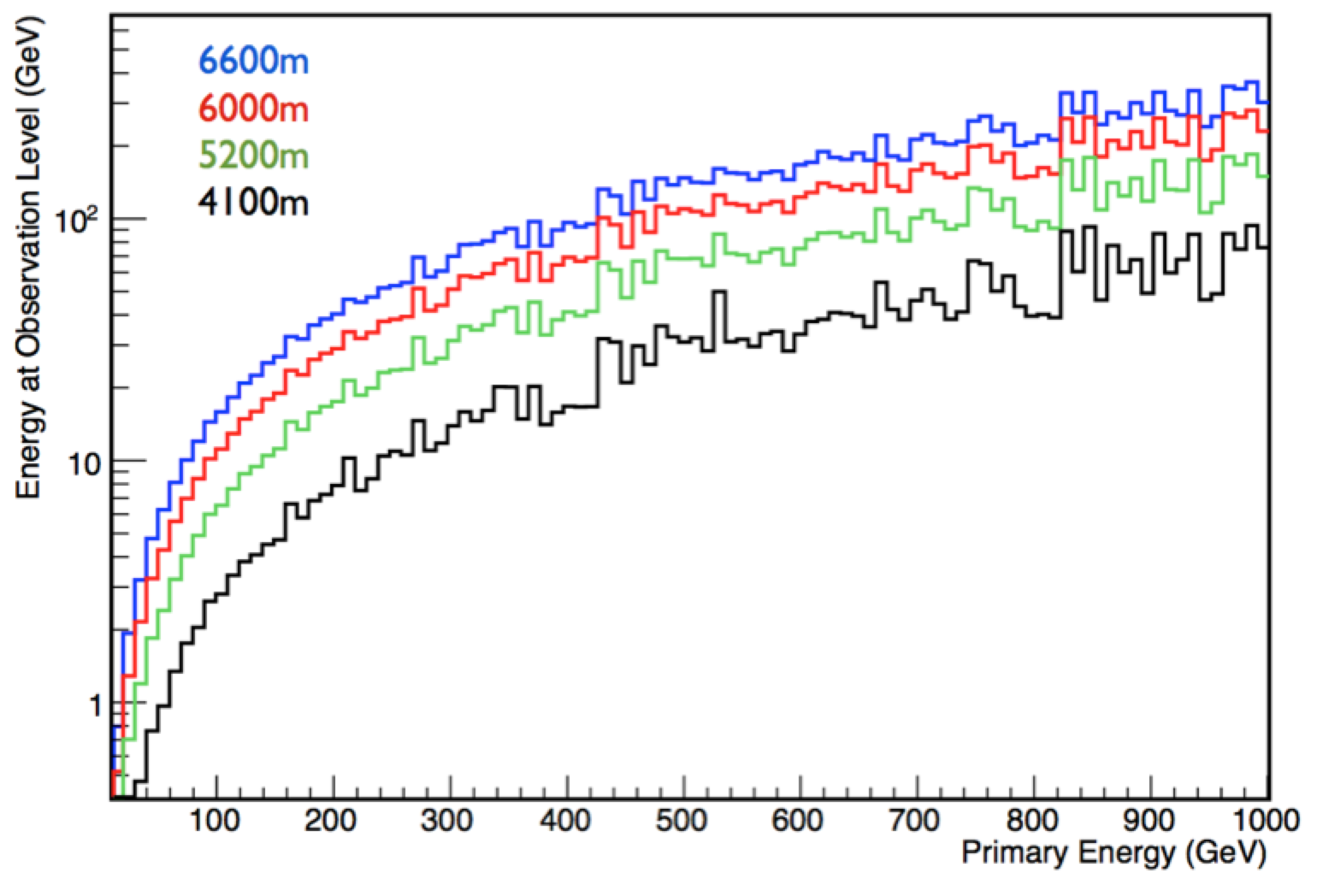}
        \caption{Surviving energy below 1 TeV.} 
    \label{fig:surviving_energy}
\end{figure}
\begin{figure}[h]
  \hspace{.7in}
    \includegraphics[width=.7\textwidth]{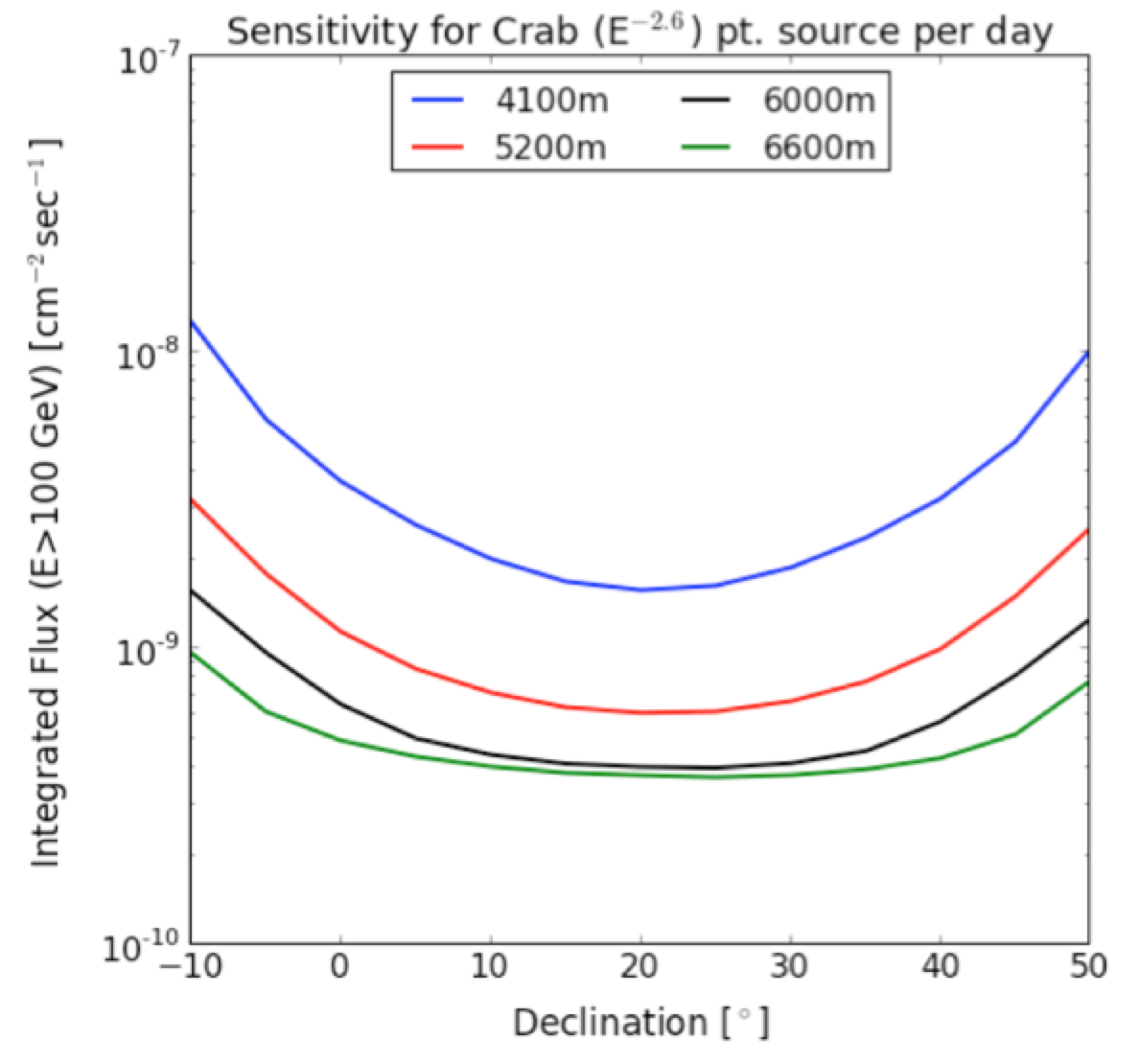}
    \caption{Sensitivity of a HAWC-like array to a gamma-ray source with the same spectrum as the Crab Nebula as a function of source declination and detector altitude.} 
    \label{fig:crab_sensitivity}
\end{figure}

Simulations of a strawman detector were performed for a variety of altitudes (4100 m for HAWC in Mexico, 5200m for plausible sites, 6000 m for maximum Southern altitudes of flat sites, and
6600 m for an extreme
reference). See Fig.~\ref{fig:surviving_energy}
for the ground level energy of gamma-rays below
1 TeV. The simulated detector was a 30$\times$30 array of 1.5 m radius and 1.5 m tall tanks separated by 3 m and with a single 10" high-quantum efficiency PMT in the water. This detector yields 
estimated Crab Sensitivities as shown in Fig.~\ref{fig:crab_sensitivity}
with the experiment sited at
20$^\circ$ South.

Any detector at higher altitudes will have to deal with higher rates. This could be ameliorated with 
a larger number of smaller tanks, local coincidence requirements, or significantly
faster trigger processing.

\subsection{Larger area}
A straightforward path to improved photon sensitivity is simply increasing the area of the detector
array. This would need to be carefully matched to the available high altitude sites, which at above 
5200m seem to be constrained to less than 20,000 m$^2$ of roughly level ground before dropping
down the slopes of the hill. Likely the best route to increased area would be via lower per-tank costs
of construction and instrumenting a couple of closely spaced hilltops with a common acquisition
system. 

\subsection{Background suppression.}
Improvements in the gamma-hadron separation directly yield improved gamma-ray rates
and decreased backgrounds. With a single detector technology (the water tanks), hadron 
rejection comes from the shower topology. That is, hadronic cascades generate sub-shower clumps
of deposited energy whereas photon showers show only the smoother electromagnetic component.
Direct detection of muons and nuclear fragments in the air showers could also be used for gamma/hadron separation. Muon detection in the water Cherenkov tank, underground muon paddles, or dividing
the tank into top (electromagnetic) and bottom (hadronic) segments are all options in this direction.
All would effectively double the cost per tank, and the trade-off between increased numbers of
tanks and increased background suppression has not yet been fully studied.

Other ideas in this direction include thin layers of liquid scintillator in thin bags within the 
tank, read out by a small 
PMT, or boron-laced plastic scintillators which observe the neutron back-splash 
from the Earth shortly after
a hadronic shower touches down. These techniques are not yet proven in large detectors, but can
also be prototyped at the HAWC Observatory in Mexico.

\subsection{Improved tank sensitivity}
Improving the sensitivity of an individual single water tank implies either increasing the light yield in the water
and/or collecting more of the photons produced in the tank. The Cherenkov light yield is difficult to
modify but adding in a scintillation component is possible with a mix of water and liquid scintillator 
(which could help prevent freezing as well). More photon collection is possible with a web of 
wavelength-shifting fibers dispersed through the water and light-piped to PMTs. With the high altitude of the potential sites, the amount of ``field-work" at the site will need to be 
kept to a minimum, so designs like this would need to be built down the mountain and taken up
largely intact.

\section{Additional technologies}
In efforts to increase either the total photon yield at the PMTs, or enhancing the discrimination 
between the background hadrons and the signal gamma-rays, ideas
such as placing scintillators above or below the tanks, adding dopants to the water,
segmenting the tanks top and bottom, or adding neutron detectors (boron-doped scintillators) under
the tanks are under consideration. Environmental concerns and the location of the HAWC site 
in Mexico, which is in a national park, have made buried detectors or dopants rather difficult to
test with the existing detectors. At the HAWC site, the installation of prototype electronics and 
different technology detectors
is relatively easy and their operation in concert with the main detector can be used for 
cross-calibration.

To more adequately address the testing of prototype detectors and novel detector technologies, 
two tanks near the counting house will be outfitted with more general readout electronics, 
and have time-tagging which is compatible with the HAWC time stamps but also easily available
to new detectors. IceCube electronics, specially modified to handle the high data rates at the site, will be the first electronics tested in these tanks. A flexible interface standard is being designed to allow
for easier access to interface with the HAWC data flow.

\section{Schedule and prospects}
Outrigger electronics development is underway at the University of Wisconsin during 2015 with
a goal of a deployable sample readout system available Spring 2016 for one of the water tanks
at HAWC in Mexico. Large-scale deployments might be possible later in 2016. Parallel work 
taking advantage of the IceCube distributed digitization electronics is also
underway with a similar prototyping time scale. Visits to potential HAWC South sites are also
ongoing. With the immense potential synergies between a Southern Hemisphere large area
water Cherenkov detector and the planned Southern CTA telescopes, further cooperation between
the groups is foreseen. 

\section{Conclusions}
The HAWC Observatory is in routine operations now, with data flowing to collaborators in the
USA and Mexico and early publications are appearing (see \cite{hawc} for an overview of 
current results). 
Design work is ongoing for a new set of electronics and smaller water tank
setups for the outriggers intended to go around the inner core of detectors at HAWC. 
This design work 
should also help to bridge the gap between the time-over-threshold electronics located in a
central counting house in HAWC (and partially inherited from the Milagro experiment)
and distributed waveform digitization with local pre-analysis
in a future HAWC-South design. Custom electronics are required for this due to the very high singles
rate at HAWC altitudes. 

There are plausible southern hemisphere sites for a detector
that would be well-matched as a ``finder-scope'' paired to the CTA sensitivity and directionality. The 
sites are in South America which could work well with CTA potential sites.
Alternate detector technologies to replace or augment the water Cherenkov technique 
are also being investigated, but would need to be sufficiently
low risk to become baseline plans.

\end{document}